\let\includefigures=\iftrue   
\input harvmac 


\input epsf 

\newcount\figno 
\figno=0 
\def\fig#1#2#3{ 
\par\begingroup\parindent=0pt\leftskip=1cm\rightskip=1cm\parindent=0pt 
\baselineskip=11pt 
\global\advance\figno by 1 
\midinsert 
\epsfxsize=#3 
\centerline{\epsfbox{#2}} 
\vskip 12pt 
{\bf Figure \the\figno:} #1\par 
\endinsert\endgroup\par } 
\def\figlabel#1{\xdef#1{\the\figno}} 


\noblackbox 
\def\IZ{\relax\ifmmode\mathchoice 
{\hbox{\cmss Z\kern-.4em Z}}{\hbox{\cmss Z\kern-.4em Z}} 
{\lower.9pt\hbox{\cmsss Z\kern-.4em Z}} {\lower1.2pt\hbox{\cmsss 
Z\kern-.4em Z}}\else{\cmss Z\kern-.4em Z}\fi}



\def\d{\partial}

%
%

%
\catcode`\@=11   
\def\slash#1{\mathord{\mathpalette\c@ncel{#1}}}   
\overfullrule=0pt

\def\CC{{\cal C}}

\def\JJ{{\cal J}}

\def\MM{{\cal M}}   
\def\NN{{\cal N}}   
\def\OO{{\cal O}}

\def\SS{{\cal S}}   
\def\TT{{\cal T}}

\def\in{\infty}
\def\go{\rightarrow}
\def\co{coordinate}

\catcode`\@=12   
   


\def\myTitle#1#2{\nopagenumbers\abstractfont\hsize=\hstitle\rightline{#1}%
\vskip 0.5in\centerline{\titlefont #2}\abstractfont\vskip .5in\pageno=0} 

\myTitle{\vbox{\baselineskip12pt\hbox{} 
\hbox{RI-12-03} 
\hbox{ITFA-2003-61}
}} {\vbox{ 
        \centerline{Comment on Counting Black Hole Microstates}\medskip
        \centerline{Using String Dualities}\medskip
        \medskip 
        }} 
\medskip 
\centerline{Assaf Shomer 
\foot{E-mail : {\tt shomer@science.uva.nl \ ;\ shomer@cc.huji.ac.il.}}
} 

\medskip 
\centerline{Racah Institute of Physics, The Hebrew University, 
Jerusalem 91904, Israel} 

\centerline{and}

\centerline{Institute of Theoretical Physics, University of Amsterdam}
\centerline{Valckenierstraat 65, 1018 XE Amsterdam, The 
Netherlands\foot{visiting.}.} 
\bigskip 
\bigskip 
\bigskip 
\noindent 

We discuss a previous attempt at a microscopic counting of the entropy of 
asymptotically flat non-extremal black-holes. This method used string dualities 
to relate 4 and 5 dimensional black holes to the BTZ black hole. 
We show how the dualities can be justified in a certain limit, equivalent to a 
near horizon limit, but the resulting spacetime is no longer asymptotically 
flat.

\Date{December 2003} 

\baselineskip=16pt

\lref\juan{J.~Maldacena, 
Adv.\ Theor.\ Math.\ Phys.\  {\bf 2} (1998) 231 ; 
Int.\ J.\ Theor.\ Phys.\  {\bf 38} (1998) 1113 ; hep-th/9711200.} 

\lref\shiftwo{

G.~T.~Horowitz and D.~L.~Welch,
Phys.\ Rev.\ Lett.\  {\bf 71}, 328 (1993)
[hep-th/9302126].

E.~Alvarez, L.~Alvarez-Gaume, J.~L.~F.~Barbon and Y.~Lozano,
Nucl.\ Phys.\ B {\bf 415}, 71 (1994)
[hep-th/9309039].

I.~Bakas,
Phys.\ Lett.\ B {\bf 343}, 103 (1995)
[hep-th/9410104].

E.~Bergshoeff and K.~Behrndt,
Class.\ Quant.\ Grav.\  {\bf 15}, 1801 (1998)
[hep-th/9803090].

E.~Cremmer, I.~V.~Lavrinenko, H.~Lu, C.~N.~Pope, K.~S.~Stelle and T.~A.~Tran,
Nucl.\ Phys.\ B {\bf 534}, 40 (1998)
[hep-th/9803259].
}

\lref\shift{
S.~Hyun,
J.\ Korean Phys.\ Soc.\  {\bf 33}, S532 (1998)
[hep-th/9704005].

H.~J.~Boonstra, B.~Peeters and K.~Skenderis,
Phys.\ Lett.\ B {\bf 411}, 59 (1997)
[hep-th/9706192].

H.~J.~Boonstra, B.~Peeters and K.~Skenderis,
Fortsch.\ Phys.\  {\bf 47}, 109 (1999)
[hep-th/9801076].
}

\lref\HorowitzWT{
G.~T.~Horowitz and D.~L.~Welch,
Phys.\ Rev.\ D {\bf 49}, 590 (1994)
[hep-th/9308077].}

\lref\StromingerSH{
A.~Strominger and C.~Vafa,
Phys.\ Lett.\ B {\bf 379}, 99 (1996)
[hep-th/9601029].}

\lref\SkenderisBS{
K.~Skenderis,
Lect.\ Notes Phys.\  {\bf 541}, 325 (2000)
[hep-th/9901050].}

\lref\SfetsosXS{
K.~Sfetsos and K.~Skenderis,
Nucl.\ Phys.\ B {\bf 517}, 179 (1998)
[hep-th/9711138].}

\lref\BanadosWN{
M.~Banados, C.~Teitelboim and J.~Zanelli,
Phys.\ Rev.\ Lett.\  {\bf 69}, 1849 (1992)
[hep-th/9204099].
}

\lref\review{
A.~W.~Peet,
hep-th/0008241.

J.~M.~Maldacena,
hep-th/9607235.
}

\lref\SeibergAD{
N.~Seiberg,
Phys.\ Rev.\ Lett.\  {\bf 79}, 3577 (1997)
[hep-th/9710009].
}

\lref\MaldacenaBW{
J.~M.~Maldacena and A.~Strominger,
JHEP {\bf 9812}, 005 (1998)
[hep-th/9804085].
}

\lref\StromingerYG{
A.~Strominger,
JHEP {\bf 9901}, 007 (1999)
[hep-th/9809027].
}

\lref\CallanDV{
C.~G.~Callan and J.~M.~Maldacena,
Nucl.\ Phys.\ B {\bf 472}, 591 (1996)
[hep-th/9602043].
}

\lref\StromingerEQ{
A.~Strominger,
JHEP {\bf 9802}, 009 (1998)
[hep-th/9712251].
}

\lref\Carlip{
S.~Carlip,
Phys.\ Rev.\ D {\bf 51}, 632 (1995)
[gr-qc/9409052].
S.~Carlip,
Phys.\ Rev.\ D {\bf 55}, 878 (1997)
[gr-qc/9606043].
}

\lref\matrx{
W.~Taylor,
Rev.\ Mod.\ Phys.\  {\bf 73}, 419 (2001)
[arXiv:hep-th/0101126].

A.~Shomer,
Phys.\ Rev.\ D {\bf 68}, 086002 (2003)
[hep-th/0303055].
}

\lref\CardyIE{
J.~L.~Cardy,
Nucl.\ Phys.\ B {\bf 270}, 186 (1986).
}

\lref\nonext{
G.~T.~Horowitz and A.~Strominger,
Phys.\ Rev.\ Lett.\  {\bf 77}, 2368 (1996)
[arXiv:hep-th/9602051].

G.~T.~Horowitz, J.~M.~Maldacena and A.~Strominger,
Phys.\ Lett.\ B {\bf 383}, 151 (1996)
[arXiv:hep-th/9603109].
}

\lref\relat{
V.~Balasubramanian and F.~Larsen,
Nucl.\ Phys.\ B {\bf 528}, 229 (1998)
[arXiv:hep-th/9802198].

Y.~Satoh,
Phys.\ Rev.\ D {\bf 59}, 084010 (1999)
[arXiv:hep-th/9810135].

M.~Cvetic and F.~Larsen,
Nucl.\ Phys.\ B {\bf 531}, 239 (1998)
[arXiv:hep-th/9805097].
}

\lref\MaldacenaUZ{
J.~M.~Maldacena, J.~Michelson and A.~Strominger,
JHEP {\bf 9902}, 011 (1999)
[arXiv:hep-th/9812073].
}

\newsec{Introduction.}

One of the celebrated successes of string theory is the Strominger-Vafa 
\StromingerSH\ microscopic entropy counting for extremal 5 dimensional 
asymptotically flat black-holes (see \review\ for review and references.). The 
same counting technique was successfully extended \CallanDV\nonext\ also to some 
non-extremal black-holes, but the reason for the successful comparison in those 
cases is less clear due to open-string strong coupling effects. A different 
approach to this problem was presented in \SfetsosXS\ where it was argued that 4 
and 5 dimensional non-extremal black-holes can be related by symmetries of 
string theory to the 3 dimensional non-extremal BTZ black hole (for related 
works see \relat). The transformations did not change the horizon area so one 
could hope that, while indirect, this method reduces the counting problem to the 
BTZ case where we have a better control over the microstates (using e.g. 
$AdS_3/CFT_2$ \juan\StromingerEQ\ or Carlip's approach \Carlip\foot{This 
approach 
is somewhat controversial due to the inclusion of negative norm states in the 
entropy counting and the choice of boundary conditions on the horizon.}). 

However, as was already noticed \SfetsosXS\ (see e.g. section 4.2.2 of 
\SkenderisBS\ for a discussion) this line of argumentation suffers from a 
caveat. 
It crucially relies on a combination of T-duality and a \co\ transformation, 
referred to as the ``shift" transformation, which effectively replaced the 
black-hole with its near horizon \shift\ (see also \shiftwo).
Closer inspection reveals that this involves a T-duality along a non-compact 
isometry. Moreover, the corresponding Killing vector was null at spatial 
infinity. This operation relates different solutions of the low energy 
supergravity but it is not entirely clear if such a transformation is a 
symmetry of string theory. Therefore, 
one could not argue for an equivalence between the initial and final 
configurations and the matching of the thermodynamic quantities was left as a 
suggestive indication that the two are somehow related.
In this note we present a way of closing this loophole, so that all the duality 
transformations are well defined. This forces the introduction of a certain 
limiting procedure which appears different but turns out to be equivalent to the 
near horizon limit used by \juan\MaldacenaBW. This explains the matching of the 
thermodynamic quantities but also shows that the counting is really done for a 
black-hole which is not asymptotically flat. We chose to focus on the 5 
dimensional case, but the result is easily extendible to other cases.

The structure of this note is as follows. In section 2 we review the``shift" 
transformation in a simple setting, discuss its shortcomings and show how one 
can replace it by a limit. In section 3 we apply this to the non-extremal 
$D1-D5-$Wave in type IIB string theory and get the BTZ$\times S^3\times T^4$ 
background. In section 4 we discuss the nature of the limit and its relation 
to the near horizon limit. In section 5 we discuss the reduction to 5 
dimensions. A short summary appears on section 6.

\newsec{The Shift Transformation}

In this section we quickly review the simplest case of the ``Shift" 
transformation \shift\ for the fundamental string solution. We then argue that a 
possible way of extending it to a symmetry of string backgrounds is via a 
certain limiting procedure.

We start with the non-extremal black string solution of type II supergravity in 
the string frame\foot{Throughout this note $g_s$ stands for the asymptotic value 
of the dilaton and $\omega_d$ for the volume of the $d$ dimensional unit sphere. 
We work in conventions where $G_N^{(10)}=8\pi^6 g_s^2 (\alpha')^4$.}

\eqn\efone{\eqalign{ds^2&={1 \over H(r)}(-f(r)dt^2+dx^2)+({1 \over 
f(r)}dr^2+r^2d\Omega_7^2) \cr B_{tx}&={f(r) \over H(r)}\tanh\alpha \quad\quad ; 
\quad\quad e^{-2\phi}={H(r) \over g_s^2}}}
where the harmonic function and the function controlling the non extremality 
are
\eqn\harm{H(r)=1+{\mu^6 \over r^6}\sinh^2\alpha \quad\quad ; \quad\quad 
f(r)=1-{\mu^6\over r^6}.}

This solution has an inner horizons at $r=0$ and an outer horizon at $r=\mu$. 
The constant part of the antisymmetric tensor $B$ is fixed so that it vanishes 
on the outer horizon as required by regularity \HorowitzWT. The coordinate $x$ 
is periodic with period $R$

\eqn\ar{x \sim x+ 2\pi R.}

The entropy is

\eqn\entro{\SS={2\pi R \mu^7 \omega_7\over 4 G_N^{(10)} }  \cosh\alpha} 

The idea behind the shift transformation is to go to the near horizon geometry 
(effectively to ``drop the $1$" from the harmonic function) using a chain of 
U-duality and coordinate transformations (assumed to be symmetry operations) in 
a way that leaves the entropy invariant.

Starting with a T-duality along $x$ we get

\eqn\ontee{\eqalign{ds^2&=-f({dt \over \cosh\alpha}-\sinh\alpha\ 
dx)^2+\cosh^2\alpha\ dx^2 + {1 \over f}dr^2+r^2d\Omega_7^2 \cr B&=0 \quad\quad ; 
\quad\quad e^{-2\phi}={R^2 \over g_s^2} .}}

Now perform the following $SL(2,R)$ change of variables. One uses an 
$SL(2,R)$ transformation in order not to change the area of the horizon and thus 
the entropy of the solution.

\eqn\sltar{\tilde{t}\equiv {1 \over \cosh\alpha} t + e^{-\alpha} x \quad ;\quad
\tilde{x}\equiv (\cosh\alpha) x.}

In terms of the new (tilded) variables the solution takes the form

\eqn\twotee{\eqalign{ds^2&=-f(d\tilde{t} -d\tilde{x})^2+ d\tilde{x}^2 + {1 \over 
f}dr^2+r^2d\Omega_7^2 \cr B&=0 \quad\quad ; \quad\quad e^{-2\phi}={R^2 \over 
g_s^2} .}}

T-dualizing back along $\tilde{x}$ gives:

\eqn\newfone{\eqalign{ds^2&={1 \over 1-f}(-fd\tilde{t}^2+d\tilde{x}^2)+({1 \over 
f}dr^2+r^2d\Omega_7^2) \cr B_{\tilde{t}\tilde{x}}&={f \over 1-f} \quad\quad ; 
\quad\quad e^{-2\phi}={1-f \over g_s^2} .}}

Comparing \newfone\ with \efone\ we see the chain of transformation amounts to
replacing $H(r)\go 1-f(r)={\mu^6 \over r^6}$ namely dropping the 1 in the 
harmonic function in \harm\ and rescaling by $\sinh^2\alpha$.
Since both T \HorowitzWT\ and S duality transformations do not change the area 
of the horizon and since \sltar\ also leaves it invariant one can argue that the 
above duality chain leaves the entropy invariant ({\it isentropic}) \SfetsosXS. 
This technique was extended 
\SfetsosXS\ to relate 4 and 5 dimensional black holes with the BTZ black hole 
where one can then use various techniques for microscopic entropy counting (see 
\SkenderisBS\ for a review.) 

However, this strategy, while very suggestive, has a loophole. After the 
coordinate transformation \sltar\ the orbit of the killing vector 
$\partial_{\tilde{x}}$ is non-compact\foot{This is a little confusing because 
naively $\tilde{x}$ parameterizes a rescaled circle. However, if the orbits of 
$\partial_{\tilde{x}}$ were compact then, starting from a point, by just going 
along the orbit of $\partial_{\tilde{x}}$ keeping $\tilde{t}$ fixed we must 
reach another point, identified with the original under \ar. But keeping 
$\tilde{t}$ fixed we move along the non-compact orbit of the \co\ $t$ which 
never returns to the original point.}. Instead \ar\ induces 
the following identification

\eqn\aride{\tilde{x} \sim \tilde{x}+2\pi{\cosh\alpha\over R} 
\quad\quad\wedge\quad\quad\ \tilde{t} \sim\tilde{t}+2\pi{e^{-\alpha}\over R}}

Therefore the T-duality transformation from \twotee\ to \newfone\ is not a 
symmetry. Moreover, the norm of the killing vector is 
$|\partial_{\tilde{x}}|^2={\mu^6 \over r^6}$ and so this isometry 
(while spacelike at any finite distance from the singularity) becomes null at 
spatial infinity\foot{This is similar to the case of T-dualizing with respect to 
the angular isometry of the plane in polar \co.}. T-duality with respect to 
isometries that are not everywhere spacelike is currently less well understood.

The main point of this comment is to suggest a strategy for closing this 
loophole. We can make $\tilde{x}$ compact while retaining a non-singular 
solution by taking the following limit

\eqn\leem{\eqalign{&R\ ,\ \alpha\ ,\ g_s\go\in \cr &\mu\ ,\ r\ ,\ {R\over 
\cosh\alpha}\equiv\tilde{R}\ ,\ {g_s\over\cosh\alpha}\equiv\tilde{g_s}\quad\sim 
fixed.}}
In this limit we get 

\eqn\dritee{\eqalign{ds^2&=-f(d\tilde{t} -d\tilde{x})^2+ d\tilde{x}^2 + {1 \over 
f}dr^2+r^2d\Omega_7^2 \cr B&=0 \quad\quad ; \quad\quad e^{-2\phi}={\tilde{R}^2 
\over \tilde{g}_s^2} ,}} which is very similar to \twotee\ but now $R\go 
\tilde{R}$, $g_s\go\tilde{g}_s$ and most importantly 
$\tilde{x}\sim\tilde{x}+2\pi \tilde{R}$, namely, $\tilde{x}$ parameterizes a 
circle of radius $\tilde{R}$.
Now we can safely T-dualize back along $\partial_{\tilde{x}}$ to get

\eqn\drifone{\eqalign{ds^2&={1 \over 1-f}(-fd\tilde{t}^2+d\tilde{x}^2)+({1 \over 
f}dr^2+r^2d\Omega_7^2) \cr B_{\tilde{t}\tilde{x}}&={f \over 1-f} \quad\quad ; 
\quad\quad e^{-2\phi}={1-f \over \tilde{g}_s^2}.}}

Comparing with \efone\ we see the effect of dropping the 1 from the harmonic 
function as before but now all the duality transformations are symmetries of 
string theory. Of course, there is a price to be paid because we took a limit 
and so the actual statement of duality will apply only to a limiting set of 
configurations.

We now notice that one can reach the endpoint \drifone\ by directly taking the 
limit \leem\ in \efone. 
When $\alpha\gg 1$ we can write $\Omega\equiv\sinh\alpha\sim\cosh\alpha\go\in$ 
so $H(r)\sim (1-f)\Omega^2$ and asymptotically \efone\ becomes

\eqn\foneinl{\eqalign{ds^2 =& {1 \over 
(1-f)\Omega^2}(-fdt^2+R^2d\theta^2)+{1\over f}dr^2+r^2d\Omega^2_7+\OO({1\over 
\Omega^2})\go\cr& {1 \over (1-f)}(-fd\tilde{t}^2+\tilde{R}^2d\theta^2)+{1\over 
f}dr^2+r^2d\Omega^2_7}}
where we rescaled the time \co\ 

\eqn\scalte{\tilde{t}\equiv {t \over \Omega}} in a manner resembling \sltar. 
Note that now the angle $\theta$ always parameterizes a circle. The B field and 
the dilaton also agree in this limit with \newfone. 

The entropy of \drifone\ is
\eqn\finent{\tilde{\SS}={A_8\over 4 G_N^{(10)}}={\tilde{R}\mu^7\omega_7 \over 
16\pi^5 \tilde{g}_s^2 (\alpha')^4}} which exactly equals \entro. At any finite 
stage during the limit the entropies agree only up to terms that 
vanish exponentially with $\alpha$\ due to the subleading term in \foneinl\ but 
the limiting configuration \drifone\ has the same entropy as \efone.

To summarize, the duality chain connecting \efone\ and \newfone\ can be 
corrected to include only symmetry operations if one takes a 
certain limit. We then ``forget" about the duality chain itself and define a 
limiting procedure that brings us directly from \efone\ to \newfone. The 
limiting configuration is the near horizon limit of the original configuration 
up to several rescalings of parameters. Both configurations have the same 
entropy. 
The specific example discussed above was used to explain our procedure but in 
itself suffers from some problems. For instance one needs to change to a weakly 
coupled description after the limit. We will not discuss this system any 
further in this note but rather apply the lesson in the D1-D5 system.

\newsec{The D1-D5 system and 5 dimensional Black Holes}

We now extend the discussion of the last section to the analysis of \SfetsosXS\ 
connecting 5 dimensional black holes with the BTZ black hole.
Start from the non-extremal $D1-D5$ system where the world-volume of the $D1$ 
brane is wrapped on a circle of radius $R$ and carries some units of momentum 
along this circle. The remaining 4 directions in the $D5$ 
world-volume are compactified on a torus with Radii $R_1,R_2,R_3,R_4$. We write 
the solution in the string frame\foot{We denote throughout the dimensionless 
metric by $d\sigma$ and the dimensionful metric by $ds$.} 

\eqn\dodf{\eqalign{\alpha' d\sigma^2&={1\over \sqrt{H_1(r)H_5(r)}} \left( 
-{f(r)\over K(r)}dt^2+K(r)(Rd\theta+{1-f(r)\over 
K(r)}\sinh\alpha_k\cosh\alpha_kdt)^2 \right)\cr &+\sqrt{{H_1(r)\over 
H_5(r)}}(R_1^2d\psi_1^2+\dots+R_4^2d\psi_4^2)+\sqrt{H_1(r)H_5(r)}({1\over 
f(r)}dr^2+r^2d\Omega_3^2) \cr 
&\CC_{t\theta}^{(2)}={1\over g_s}{R\over\alpha'}{f(r)\over H_1(r)}\tanh\alpha_1 
\quad\quad ; \quad 
 e^{-2\phi}={H_5(r)\over g_s^2H_1(r)}\cr
 &\CC_{t\theta\psi_1\dots\psi_4}^{(6)}={v\over g_s}{R\over\alpha'}{f(r)\over 
H_5(r)}\tanh\alpha_5.}}

where 

\eqn\ufde{\eqalign{H_{1/5}(r)&=1+{\mu^2\over r^2}\sinh^2\alpha_{1/5} \ ;\ 
K(r)=1+{\mu^2\over r^2}\sinh^2\alpha_k\cr f(r)&=1-{\mu^2\over r^2}\ ;\
v={V\over (\alpha')^2}={R_1R_2R_3R_4\over(\alpha')^2}}} and the rest of the IIB 
supergravity fields vanish. 

This configuration has $3$ conserved charges 

\eqn\charges{\eqalign{\NN_1 &={v\mu^2\over g_s\alpha'} 
\sinh\alpha_1\cosh\alpha_1 ,\cr\NN_5 &={\mu^2\over 
g_s\alpha'}\sinh\alpha_5\cosh\alpha_5,\cr
\NN_k &={R^2v\mu^2\over g_s^2(\alpha')^2}\sinh\alpha_k\cosh\alpha_k,}} and the 
ADM mass, entropy and temperature are given by 

\eqn\thermo{\eqalign{\MM&={R v \mu^2 \over 2g_s^2 (\alpha')^2} (\cosh 2\alpha_1 
+ \cosh 2\alpha_5  + \cosh 2\alpha_k),\cr\SS &={2\pi Rv\mu^3\over 
g_s^2(\alpha')^2}
\cosh\alpha_1\cosh\alpha_5\cosh\alpha_k,\cr
\TT&={1\over 2\pi\mu\cosh\alpha_1\cosh\alpha_5\cosh\alpha_k}.}}

This configuration can be seen as a thermal excitation of the supersymmetric 
$D1-D5$ configuration and therefore satisfies a BPS condition  

\eqn\bpsdodf{\MM\geq {\NN_k\over R} + {R\over g_s\alpha'}(\NN_1+v\NN_5)} 
satisfied here thanks to the property $\cosh\alpha\geq\sinh\alpha$ (equality 
holding only for the asymptotic value $\alpha=\in$). 

In \SfetsosXS\ a duality chain (including T-duality along non compact orbits) 
was used to relate this solution to the BTZ black-hole \BanadosWN.
As in the previous section, we can close the loophole in the dualities for the 
price of taking a limit which in turn we can think of as an independent and 
equivalent way of performing the dualities.
The limit in this case is deduced in a similar way to \leem :

\eqn\leema{\eqalign{&\Omega_{1,5}\equiv\cosh\alpha_{1,5}\sim\sinh\alpha_{1,5} 
\go\in  \cr &\alpha'\ , R\ , R_{1,2,3,4}\go\in }} keeping the following 
quantities fixed
\eqn\leemb{\eqalign{&\tilde{\alpha'}\equiv {\alpha'\over\Omega_1\Omega_5} \ ,\
\tilde{R}\equiv {R\over\Omega_1\Omega_5}\ , \
\tilde{R_i}\equiv {R_i\over\Omega_5}\ ,\
\tilde{g_s}\equiv g_s{\Omega_1\over\Omega_5} }}

As in \scalte\ we have to rescale the time \co\  
\eqn\leemc{\tilde{t}\equiv{t\over \Omega_1\Omega_5} \Rightarrow \tilde{E}\equiv 
\Omega_1\Omega_5 E.}

As opposed to \leem\ we do not need to send $g_s\go\in$ since the dilaton in 
\dodf\ involves a ratio of the harmonic functions but we do need to send 
$\alpha'\go\in$ due to an overall $\Omega_1\Omega_5$ factor in front of the 
metric. The common feature in both cases is a rescaling of the Planck mass. Note 
also that even though $l_p\go\in$ this is a low energy limit\foot{This is like 
in 
M(atrix) theory where $\alpha'\go\in$ before scaling the Planck mass due to the 
vanishing spacelike circle \SeibergAD.} because from \leemc\ we see that 
${E\over m_p}\sim {1\over \sqrt{\Omega_1\Omega_5}}\go 0$.

Taking the limit while keeping all tilded quantities finite we get the following 
solution

\eqn\beetz{\eqalign{\tilde{\alpha}' d\sigma^2&={1\over 1-f} \left( -{f\over 
K}d\tilde{t}^2+K(\tilde{R}d\theta+{1-f\over 
K}\sinh\alpha_k\cosh\alpha_kd\tilde{t})^2 \right)\cr 
&+\tilde{R}_1^2d\psi_1^2+\dots+\tilde{R}_4^2d\psi_4^2+(1-f)({1\over 
f}dr^2+r^2d\Omega_3^2) \cr 
&\CC_{\tilde{t}\theta}^{(2)}={1\over 
\tilde{g}_s}{\tilde{R}\over\tilde{\alpha}'}{f\over 1-f} \quad\quad ; \quad 
 e^{-2\phi}={1\over \tilde{g}_s^2}\cr
 &\CC_{\tilde{t}\theta\psi_1\dots\psi_4}^{(6)}={\tilde{v}\over 
\tilde{g}_s}{\tilde{R}\over\tilde{\alpha}'}{f\over 1-f}.}}
 
Writing the $r$ dependence explicitly in the Einstein frame we get

\eqn\expar{\eqalign{\sqrt{\tilde{g}_s}\tilde{\alpha}'d\sigma_E^2&=-{{r^2\over\mu
^2}-1\over 1+{\mu^2\over r^2}\sinh^2\alpha_k} 
d\tilde{t}^2+({r^2\over\mu^2}+\sinh^2\alpha_k) \left(\tilde{R}d\theta+{\mu^2 
\sinh\alpha_k\cosh\alpha_k \over  (r^2+\mu^2\sinh^2\alpha_k)}d\tilde{t} 
\right)^2 \cr &+{\mu^2\over 
r^2-\mu^2}dr^2+\mu^2d\Omega_3^2+\tilde{R}_1^2d\psi_1^2\dots+\tilde{R}_4^2d\psi_4
^2,}} the $\tilde{t},\theta,r$ part of the solution is identified with the 
general non-extremal BTZ black-hole solution of 3d gravity with negative 
cosmological constant $\Lambda=-{1\over l^2}$ \BanadosWN\ 

\eqn\btzmet{ds^2_{BTZ}=-{(\rho^2-\rho_+^2)(\rho^2-\rho_-^2)\over 
l^2\rho^2}d\tau^2+\rho^2(d\varphi^2+{\rho_+\rho_-\over 
l\rho^2}d\tau)^2+{l^2\rho^2\over(\rho^2-\rho_+^2)(\rho^2-\rho_-^2)}d\rho^2} 
where we identified the Planck scales and used the following dictionary

\eqn\dicti{\eqalign{l_{BTZ}=\mu\quad &;\quad 
\tau_{BTZ}=\tilde{t}{\mu\over\tilde{R}}\quad ;\quad 
\rho_{BTZ}^2={\tilde{R}^2\over\mu^2}(r^2+\mu^2\sinh^2\alpha_k) \cr
\rho_+&=\tilde{R}\cosh\alpha_k\quad\quad ;\quad\quad 
\rho_-=\tilde{R}\sinh\alpha_k}}
The full solution \beetz\expar\ is thus BTZ$\times S^3\times T^4$. 

\newsec{The limit}

The charges and the thermodynamical quantities associated with \expar\beetz\ can 
be inferred from the BTZ factor to be\foot{Note that dimensionful quantities had 
to be rescaled by powers of ${\mu\over\tilde{R}}$ due to the transformation 
$\tilde{E}={\mu\over\tilde{R}}E_{BTZ}$ induced from \dicti.}

\eqn\entbtz{\eqalign{
\tilde{\MM}&={\mu\over\tilde{R}}{\rho_+^2+\rho_-^2 \over 8 l^2 
G_N^{(3)}}={\mu\over\tilde{R}}{\tilde{R}^2(\cosh^2\alpha_k+\sinh^2\alpha_k) 
(2\pi)^4\tilde{v}(2\pi^2)\mu^3\over 8\mu^2(8\pi^6 \tilde{g_s}^2 
(\tilde{\alpha}')^2)}={\tilde{R} \tilde{v}\mu^2\over 2 \tilde{g_s}^2 
(\tilde{\alpha}')^2}\cosh 2\alpha_k\cr 
\tilde{\JJ}&={2\rho_+\rho_- \over 8 l G_N^{(3)}}={\tilde{R}^2 
\tilde{v}\mu^2\over  \tilde{g_s}^2 
(\tilde{\alpha}')^2}\sinh\alpha_k\cosh\alpha_k\cr 
\tilde{\SS}&={2\pi \tilde{R}\tilde{v}\mu^3\over 
\tilde{g}_s^2(\tilde{\alpha}')^2}\cosh\alpha_k\cr
\tilde{\TT}&={\mu\over\tilde{R}}{\rho_+^2-\rho_-^2 \over 2\pi \rho_+ l^2}={1 
\over 2\pi\mu\cosh\alpha_k}}}

The entropy and temperature in \entbtz\ are exactly the limiting values  of 
those in \thermo\foot{Note that one has to rescale units of energy as in 
\leemc.}.
The charges in \charges\ are also finite in the limit with $\NN_k$ approaching 
the BTZ angular momentum charge $\tilde{\JJ}$. Only the ratio of the two 
remaining charges encodes independent information about \beetz, namely, the 
dimensionless volume\foot{This dimensionless volume $\tilde{v}$ 
is related to that of the original $T^4$ in \dodf\ through the ratio 
$\sqrt{\tilde{v}}=\sqrt{v}{\Omega_1\over\Omega_5}$ which is  
an arbitrary constant in the limit.} of the $T^4$. 

In summary 
 
\eqn\leechar{\eqalign{&\NN_1 \go{\tilde{v}\mu^2\over \tilde{g}_s\tilde{\alpha}'} 
\ ;\ \NN_5 \go{\mu^2\over \tilde{g}_s\tilde{\alpha}'}\ ;\  
{\NN_1\over\NN_5}=\tilde{v},\cr &\NN_k \go\tilde{\JJ}\ ; \ \SS \go\tilde{\SS}\ ; 
\ \TT \go\tilde{\TT}.}}

The mass, however,  diverges after we rescale the energy. This is not a problem 
since also the BPS ``zero point" energy corresponding to the supersymmetric 
solution \bpsdodf\ diverges in exactly the same way. The physical information 
resides in the fluctuations above the BPS mass which when appropriately rescaled 
as dictated by \leema\leemb\leemc\ remain finite\foot{Again this is like the 
situation in 
M(atrix) theory where the mass of the $D0$ brane diverges but the light-cone 
Hamiltonian is scaled to remain finite \SeibergAD\ (see e.g. discussion in 
\matrx).}

\eqn\leemas{\Omega_1\Omega_5\left( \MM - {R\over g_s\alpha'}(\NN_1+v\NN_5) 
\right)\longrightarrow\tilde{\MM}\geq{\tilde{\JJ}\over\tilde{R}}.}  The last 
inequality is the BPS condition for BTZ black-holes which follows in the limit 
from \bpsdodf.

Actually, although it appears different, this limit gives exactly the same 
result as the ``Near Inner Horizon Limit" (NIHL) for non-extremal branes defined 
in \juan\MaldacenaBW. This limit drops the 1 from $H_1,H_5$ by sending 
$\alpha'\go 0$ keeping fixed
\eqn\nihel{U={r\over \alpha'} \ ;\ U_0={\mu\over \alpha'} \ ;\ R \ ;\ v\ ;\ 
g_s,} which in turn keep the charges $\NN_1  \ ;\ \NN_5 \ ;\ \NN_k \ ;\ \SS  \ 
;\ \TT$ fixed\foot{Notice that this ``NIHL" necessarily involves the ``dilute 
gas 
condition" $\alpha_1,\alpha_5\go\in$.}. This is not surprising because, after 
all, the motivation behind the shift transformation was to replace (using 
dualities) the black-hole with its near horizon. It appears that to do this 
consistently one has to take the near horizon limit.

\newsec{Reducing to 5 dimensions}

When we reduce the asymptotically flat \dodf\ on $\theta,\psi_1 ,\dots,\psi_4$ 
we get a 5 dimensional non-extremal black hole. The metric is given in the 
Einstein frame by

\eqn\metfiv{\eqalign{ds_E^2&=\lambda^{-{2\over3}}(-fdt^2)+
\lambda^{{1\over3}}({1\over f}dr^2+r^2d\Omega_3^2) \cr
\lambda(r)\equiv H_1H_5K &=(1+{\mu^2\over r^2}\sinh^2\alpha_1)(1+{\mu^2\over 
r^2}\sinh^2\alpha_5)(1+{\mu^2\over r^2}\sinh^2\alpha_k).}}
Notice that the 5 dimensional Schwarzschield solution is also a member of this 
family with all charges equal to zero. The entropy of this solution is exactly 
equal to that of \dodf, namely the one in \thermo.

Taking the limit \leema\leemb\leemc\ in \metfiv\ or alternatively reducing 
\beetz\ on 
$\theta,\psi_1,\dots,\psi_4$ we get another 5 dimensional black hole with metric 

\eqn\redfiv{\eqalign{ds_E^2&=\tilde{\lambda}^{-{2
\over 3}}(-fd\tilde{t}^2)+\tilde{\lambda}^{{1\over 3}}({1\over 
f}dr^2+r^2d\Omega_3^2) \cr \tilde{\lambda}&=(1-f)^2K=({\mu\over 
r})^4(1+{\mu^2\over r^2}\sinh^2\alpha_k) }}
whose entropy is equal to that of \beetz\ in \entbtz.

The condition for validity of the dimensional reduction are that one focuses on 
excitations that are well below the threshold for KK and winding modes. Since in 
the limit \leema\leemb\leemc\ we keep the tilded quantities fixed and send 
$\Omega_{1,5}\go\in$ the only non-trivial condition is to have in the original 
$D1-D5$ system \dodf\ $RE\ll 1$\foot{This regime seems related to the one 
studied in \StromingerYG\MaldacenaUZ.}.

\newsec{Summary and Discussion}

The main point in \SfetsosXS\ was to use the shift transformation to relate 
\beetz\ with \metfiv. The thermodynamical quantities seemed to agree but since 
this transformation is not a symmetry one could not claim an equivalence. The 
path we have chosen here suggests instead to relate \beetz\ and the limit of 
\metfiv, namely \redfiv. 
Focusing on the low energy excitations of the non-extremal $D1-D5$ system 
\dodf\ satisfying $RE\ll 1$ one can take a limit which gives the general 
non-extremal BTZ$\times S^3 \times T^4$ background \beetz\ and which at the same 
time is well approximated by the dimensional reduction to the non-extremal 
(albeit not most general) 5 dimensional black hole \redfiv. Unfortunately, 
although \redfiv\ is a limiting 
configuration in the family \metfiv\ of asymptotically flat black-holes, it is 
not asymptotically flat itself, but rather asymptotes to a space that is 
conformal to $AdS_2\times S^3$. This is a reflection of the fact that \beetz\ is 
the NHL of \dodf\ and as such is not asymptotically flat but rather 
asymptotically $AdS_3$. The relation between \beetz\ and \redfiv\ is not a 
duality but dimensional reduction along the angular isometry $\d_{\theta}$, just 
like \metfiv\ is a dimensional reduction of \dodf. 

Nevertheless since the entropy of the configuration \metfiv\ is equal to that of 
\redfiv\ which is the result of the limit and since our limiting procedure 
itself is essentially a near-horizon limit it is plausible to argue that the 
entropy counting done in \redfiv\ or \beetz\ using entropy counting techniques 
for the BTZ black hole is closely related to the counting of degrees of freedom 
in the original asymptotically flat configuration.

\bigskip
\bigskip
\bigskip
\bigskip
\bigskip
\bigskip
\bigskip
\bigskip

\centerline{\bf Acknowledgments} 

We like to thank Riccardo Argurio, Jan de Boer, Miranda Cheng, Emiliano 
Imeroni, Asad Naqvi and Jan Troost for discussions and comments on the 
manuscript. Special thanks to Kostas Skenderis for pointing out \shift\ as 
well as for many discussions and carefully reading the manuscript. 
Thanks also to Liat Maoz for a clear introduction to 3d gravity. This work is 
supported by a Clore fellowship.

\listrefs   
   
\end   
   
   
\newsec{Personal Appendix}

\subsec{where are all the $\pi s$ and the $2 s$}

\eqn\ent{\SS={A^{(3)}\over 4G_N^{(3)}}={2\pi\rho_+ \times 
(2\pi)^4v\times2\pi^2\mu^3 \over 4\times 8\pi^6 
\tilde{g_s}^2(\tilde{\alpha}')^2}={2\pi\rho_+v\mu^3\over 
\tilde{g_s}^2(\tilde{\alpha}')^2}}
where we used that $\omega_3=2\pi^2$ and $G_N^{(10)}=8\pi^6 \tilde{g_s}^2 
(\tilde{\alpha}')^4$.

\subsec{why \dodf\ is the same as $(7.1)$ in Peet or $(5.27)$ in MAGOO}

for that we need to show why 

\eqn\petmaone{-1+(1-f)\cosh^2\alpha_k={1\over 
1+(1-f)\sinh^2\alpha_k}(-f+(1-f)^2\sinh^2\alpha_k\cosh^2\alpha_k).}

multiplying the L.H.S by the denominator 

\eqn\petmatwo{\eqalign{&(-1+(1-f)\cosh^2\alpha_k)(1+(1-f)\sinh^2\alpha_k)=\cr 
&-1+(1-f)(\cosh^2\alpha_k-\sinh^2\alpha_k)+(1-f)^2\sinh^2\alpha_k\cosh^2\alpha_k
=\cr &-f+(1-f)^2\sinh^2\alpha_k\cosh^2\alpha_k.}}

\subsec{equation \redfiv}

there is a diverging factor in front of the metric from the fact that in the 
limit \leema\leemb\leemc\ $\lambda\sim\tilde{\lambda}\Omega_1\Omega_5$ but 
theres also the 
same diverging factor in the Planc scale 
$G_N^{(5)}\sim\tilde{G}_N^{(5)}\Omega_1\Omega_5$ with 
$G_N^{(5)}=(l_p^{(5)})^3={g_s^2(\alpha')^2\over Rv}$

\subsec{Conditions for dimenional reduction}

We start with the KK modes

\eqn\condimr{E\ll{1\over R},{1\over R_i},{1\over \mu}.}

The way the relevant dimensionless ratios scale in our limit is

\eqn\scldi{RE=\tilde{R}\tilde{E}\ ;\ R_iE=\tilde{R_i}\tilde{E}{1\over \Omega_1} 
\ ;\ \mu E=\mu\tilde{E}{1\over \Omega_1\Omega_5}.}

For the winding modes we need only (because the sphere is simply connected)

\eqn\winco{E\ll RT, R_iT} with $T={1\over 2\pi\alpha'}$ the string tension.

The dimensionless ratios scale as

\eqn\sclwin{{E\over RT}={\tilde{E}\over\tilde{R} \tilde{T}}{1\over 
\Omega_1\Omega_5}\ ;\ {E\over R_iT}={\tilde{E}\over\tilde{R_i} \tilde{T}}{1\over 
\Omega_5}.}